\documentclass[aps,prl,twocolumn,groupedaddress]{revtex4-1}
\usepackage{graphicx}
\usepackage{amssymb}
\usepackage{amsmath}

\usepackage{color}

\begin{document}

%\preprint{}

\title{{\em Ab initio} many-body calculations of the $^3$H($d$,$n$)$^4$He and $^3$He($d$,$p$)$^4$He fusion}


%\thanks{}
\author{Petr Navr{\'a}til$^{1,2}$ and Sofia Quaglioni$^2$}
%%\email[]{navratil@triumf.ca}
\affiliation{$^1$TRIUMF, 4004 Wesbrook Mall, Vancouver, BC V6T 2A3, Canada\\
$^2$Lawrence Livermore National Laboratory, P.O. Box 808, L-414, Livermore, CA 94551, USA}
%\author{Sofia Quaglioni}
%%\email[]{quaglioni1@llnl.gov}
%\affiliation{Lawrence Livermore National Laboratory, P.O. Box 808, L-414, Livermore, CA 94551, USA}

\date{\today}
\begin{abstract}
We apply the {\em ab initio} no-core shell model/resonating group method approach to calculate the cross sections of the $^3$H($d$,$n$)$^4$He and $^3$He($d$,$p$)$^4$He fusion reactions. These are important reactions for the Big Bang nucleosynthesis and the future of energy generation on Earth. Starting from a selected similarity-transformed chiral nucleon-nucleon interaction that accurately describes two-nucleon data, we performed many-body calculations that predict the S-factor of both reactions. Virtual three-body breakup effects are obtained by including excited pseudostates of the deuteron in the calculation. Our results are in satisfactory agreement with experimental data and pave the way for microscopic investigations of polarization and electron screening effects, of the $^3$H($d$,$\gamma$)$^5$He radiative capture and other reactions relevant to fusion research.
\end{abstract}

%\keywords{}

\maketitle

The $^3$H($d$,$n$)$^4$He and $^3$He($d$,$p$)$^4$He reactions are leading processes in the primordial formation of the very light elements (mass number, $A\le7$), affecting the predictions of Big Bang nuleosynthesis for light nucleus abundances~\cite{JCAP12}. With its low activation energy and high yield, $^3$H($d$,$n$)$^4$He is also the easiest reaction to achieve on Earth, and is pursued by research facilities directed toward developing fusion power by either magnetic ({\em e.g.}\ ITER~\cite{iter}) or inertial ({\em e.g.}\ NIF~\cite{nif}) confinement.   
The cross section for the $d+^3$H fusion is well known experimentally%~\cite{dT}
, while more uncertain is the situation for the branch of this reaction, $^3$H$(d, \gamma)^5$He, that %produces $17.9$ MeV $\gamma$-rays~\cite{dTgamma} and 
is being considered as a possible plasma diagnostics in modern fusion experiments~\cite{NIFdiagnostic}. Larger uncertainties dominate also the $^3$He($d$,$p$)$^4$He reaction %~\cite{d3He} 
 that is known for presenting considerable 
electron-screening effects at energies accessible by beam-target experiments.  
Here, the electrons bound to the target (usually a neutral atom or molecule)  
lead to  enhanced values (increasingly with decreasing energy) for the reaction-rate, 
effectively preventing direct access to the astrophysically relevant bare-nucleus cross section. Consensus on the physics mechanism behind this enhancement is not been reached yet~\cite{chaos}, largely because of the difficulty of determining the absolute value of the bare cross section. 
Past  theoretical investigations of these fusion reactions include various $R$-matrix analyses of experimental data at higher energies~\cite{Hale87,Barker97,Simeckova99,Desc04}  as well as microscopic calculations with phenomenological interactions~\cite{Langanke90_91,Csoto97}. However, in view of remaining experimental challenges  (some of which described above) and the large role played by theory in extracting the astrophysically important information, it would be highly desirable to achieve a microscopic description of the $^3$H($d$,$n$)$^4$He and $^3$He($d$,$p$)$^4$He fusion reactions 
that encompasses the dynamic of all five nucleons and is based on the fundamental underlying physics: the realistic interactions among nucleons and the structure of the fusing nuclei.
 
In this Letter, we present the first {\it ab initio} many-body calculation of the $^3$H($d$,$n$)$^4$He and $^3$He($d$,$p$)$^4$He fusion reactions starting from a nucleon-nucleon ($NN$) interaction that describes two-nucleon properties with high accuracy. 
The present calculations are performed in the framework of the {\em ab initio} no-core shell model/resonating-group method (NCSM/RGM)~\cite{NCSMRGM,NCSMRGM_IT,NCSMRGM_dalpha}, %{\em ab initio} no-core shell model (NCSM)~\cite{NCSMC12} combined with the resonating-group method (RGM)~\cite{RGM,RGM1}, or  {\em ab initio} NCSM/RGM~\cite{NCSMRGM,NCSMRGM_IT,NCSMRGM_dalpha}, 
a unified approach to bound and scattering states of light nuclei.
We use, in particular, the orthonormalized %NCSM/RGM 
many-body wave functions %given by
($\nu$ being the channel index) 
\begin{eqnarray}
|\Psi^{J^\pi T}\rangle &=& \sum_{\nu} \int dr r^2 \, \hat{\mathcal A}_{\nu}|\Phi^{J^\pi T}_{\nu r}\rangle  \frac{[{\cal N}^{-1/2}\chi]_\nu(r)}{r}
\, , \label{trial}
\end{eqnarray}
with inter-cluster antisymmetrizer for the ($A{-}a$,$a$) partition $\hat{\mathcal A}_{\nu}$, center-of-mass separation $\vec r_{A-a,a}$, and binary-cluster channel states
\begin{eqnarray}
|\Phi^{J^\pi T}_{\nu r}\rangle &=& \Big [ \big ( \left|A{-}a\, \alpha_1 I_1^{\,\pi_1} T_1\right\rangle \left |a\,\alpha_2 I_2^{\,\pi_2} T_2\right\rangle\big ) ^{(s T)}\nonumber\\
&&\times\,Y_{\ell}\left(\hat r_{A-a,a}\right)\Big ]^{(J^\pi T)}\,\frac{\delta(r-r_{A-a,a})}{rr_{A-a,a}}\,.\label{basis}
\end{eqnarray}
The inter-cluster relative-motion wave functions $\chi^{J\pi T}_\nu(r)$ satisfy the integral-differential coupled-channel equations
\begin{equation}
{\sum_{\nu^\prime}\!\!\int \!\!dr^\prime r^{\prime\,2}} [{\mathcal N}^{-\frac12}{\mathcal H}\,{\mathcal N}^{-\frac12}]_{\nu\nu^\prime\,}\!(r,r^\prime)\frac{\chi_{\nu^\prime} (r^\prime)}{r^\prime} \!=\! E\,\frac{\chi_{\nu} (r)}{r}  \label{RGMeq}
\end{equation}
with bound- or scattering-state boundary conditions. 
Here ${\mathcal H}^{J^\pi T}_{\nu\nu^\prime}(r, r^\prime)$ and ${\mathcal N}^{J^\pi T}_{\nu\nu^\prime}(r, r^\prime)$, commonly referred to as integration kernels, are respectively the Hamiltonian and overlap (or norm) matrix elements over the antisymmetrized basis of Eq.~(\ref{basis}).  
They contain all nuclear structure and antisymmetrization properties of the problem. %Further details can be found in Refs.~\cite{NCSMRGM} and~\cite{NCSMRGM_dalpha}.

In the present application we investigate reactions involving $A{=}5$ nucleons, characterized by a deuteron-nucleus entrance and nucleon-nucleus exit channels [$a{=}2$ and $a{=}1$ in Eq.~(\ref{basis}), respectively]. 
The NCSM/RGM formalism for an $a{=}1$ projectile was presented in Ref.~\cite{NCSMRGM} (where the interest reader can find further details on the approach), while the deuteron projectile formalism was introduced in Ref.~\cite{NCSMRGM_dalpha}, where we investigated the $d$-$^4$He system. To calculate the $^3$H($d$,$n$)$^4$He and $^3$He($d$,$p$)$^4$He reactions, we had to address the additional contributions of matrix elements (two for the norm and 5 for the Hamiltonian kernel, respectively) between the two mass partitions: ($A{-}1$,$1$) and ($A{-}2$,$2$). These extensions of the formalism %for these contributions 
will be described in detail in a separate paper. 

The input into Eq.~(\ref{RGMeq}) are: $(i)$ the nuclear Hamiltonian, particularly the chiral N$^3$LO $NN$ potential of Ref.~\cite{N3LO}, which we soften by a similarity renormalization group (SRG) transformation~\cite{SRG,Roth_SRG} characterized by an evolution parameter $\Lambda$; and $(ii)$ the eigenstates of the interacting nuclei, {\em i.e.} $^2$H, $^3$H, $^3$He and $^4$He, calculated within the NCSM~\cite{NCSMC12}. In this first attempt of providing an {\em ab initio} description of the $d+^3$H ($d+^3$He) fusion, we omit both the chiral and SRG-induced three-nucleon ($NNN$) force components of the Hamiltonian, and select a value of the SRG parameter ($\Lambda=1.5$ fm$^{-1}$) for which we reproduce the experimental $Q$ value of both reactions within $1\%$. While a complete ($\Lambda$-independent) calculation should include these terms (and efforts in this direction are under way), we argue that this is a fair approximation for the time being. Indeed, for these very light nuclei the initial attractive chiral $NNN$ force cancels to some extent with that induced by the SRG evolution of the $NN$ potential, which is repulsive in this $\Lambda$ range~\cite{JNF09}.   
Further, even though the fusion proceeds at very low energies, the deformation and virtual breakup of the reacting nuclei cannot be disregarded, particularly for the weakly-bound deuteron. A proper treatment of deuteron-breakup effects requires the inclusion of three-body continuum states (neutron-proton-nucleus) and is very challenging. 
In this first fusion application we limit ourselves to binary-cluster channels and approximate virtual three-body breakup effects by discretizing the continuum with excited deuteron pseudostates, strategy that proved successful in our $d$-$^4$He calculations~\cite{NCSMRGM_dalpha}.

\begin{table}[tb]
\caption{Calculated g.s.\ energies and point-proton rms radii of $^2$H, $^3$H, $^3$He, and $^4$He obtained by using the SRG-N$^3$LO $NN$ potential with $\Lambda{=}1.5$ fm$^{-1}$ are compared to the corresponding experimental values. The NCSM calculations were performed in an HO space with $N_{\rm max}{=}12$ and $\hbar\Omega{=}14$ MeV.}
\begin{ruledtabular}
\begin{tabular}{lcccc}
&\multicolumn{2}{c}{$E_{\rm g.s.}$ [MeV] }&\multicolumn{2}{c}{$\langle r^2_{\rm p}\rangle^{1/2}$ [fm]}\\[0.2mm] \cline{2-3}\cline{4-5}
& Calc. & Expt.& Calc.& Expt. \\
\hline
$^{2}$H & ~~\!-2.20 & ~~\!-2.22 & 1.84& \!\hspace{-7mm}1.96\\
$^3$H & ~~\!-8.27 & ~~\!-8.48 & 1.64 & \!\hspace{-7mm}1.60 \\
$^3$He &  ~~\!-7.53 & ~~\!-7.72 & 1.81 & \!\hspace{-7mm}1.77\\
$^{4}$He & -28.22  & -28.29 & 1.49 & 1.467(13)
\end{tabular}
\end{ruledtabular}
\label{tab:gs}
\end{table}
We start by discussing our results for the ground states (g.s.)\ of $d$, $^3$H, $^3$He and $^4$He, the energies and radii of which are compared to experiment in Table~\ref{tab:gs}. 
The soft $NN$ interaction (SRG-N$^3$LO with $\Lambda{=}1.5$~fm$^{-1}$) and harmonic oscillator (HO) frequency ($\hbar\Omega{=}14$ MeV) adopted are the same as in the $d$-$^4$He study of Ref.~\cite{NCSMRGM_dalpha}. Energy convergence (at $\leq 20$~keV level) of the present results is reached for an HO basis size of $N_{\rm max}{=}12$, where we also find a weak  %(less than $2\%$ variation) 
frequency dependence in the range $11{\le}\hbar\Omega{\le}18$ MeV.  

\begin{figure}%[t]
\includegraphics*[width=0.75\columnwidth]{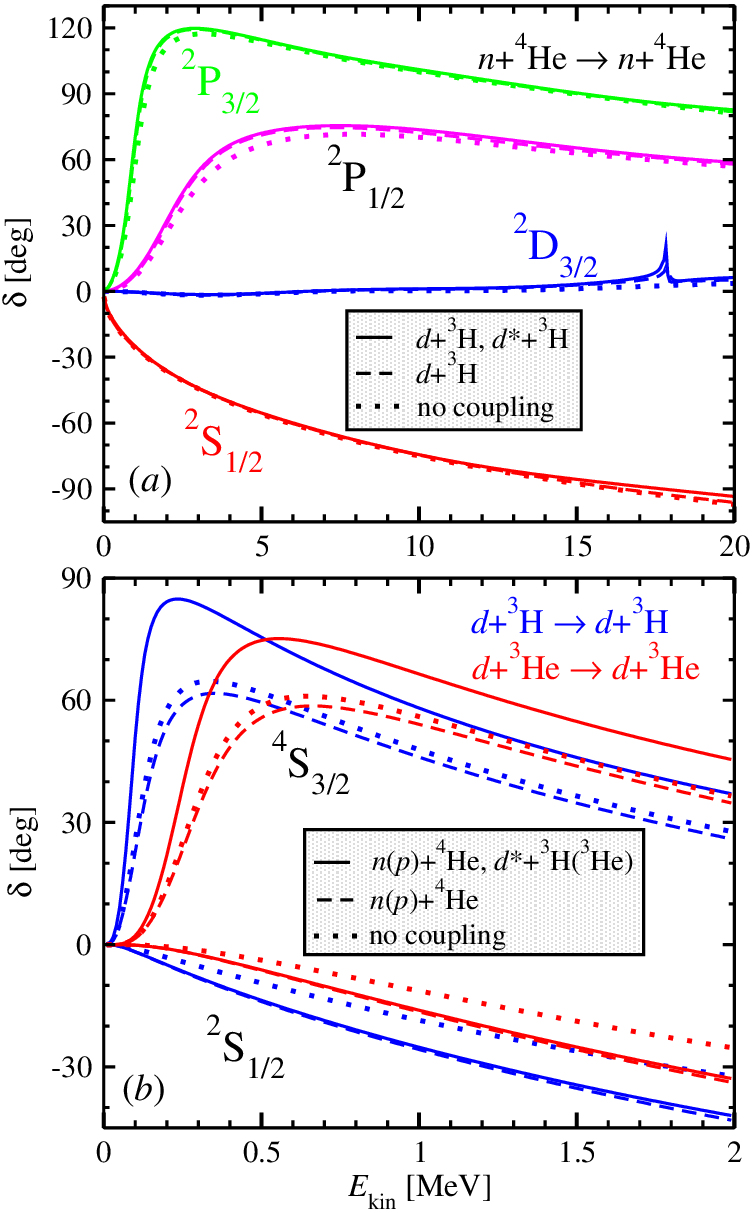}
\caption{(Color online) Calculated elastic $n$-$^4$He (a), $d$-$^3$H and $d$-$^3$He (b) phase shifts. The dashed (dotted) lines are obtained %show the results 
with (without) coupling of the $n$($p$)-$^4$He and $d$-$^3$H($^3$He) channels and all nuclei in their g.s. The full lines represent calculations that further couple channels with one $^3S_1{-}^3D_1$ deuteron pseudostate. The SRG-N$^3$LO $NN$ potential with $\Lambda{=}1.5$ fm$^{-1}$ and the HO space with $N_{\rm max}{=}12$ ($N_{\rm max}{=}13$ for the negative parity) and $\hbar\Omega{=}14$ MeV were used.}
\label{fig:nHe4_phase_shifts}
\end{figure}
Next, we consider the elastic phase shifts for both entrance and exit channels. In the past, we had already studied $n(p)$-$^4$He scattering within the NCSM/RGM~\cite{NCSMRGM,NCSMRGM_IT}.
Here, we extend those calculations by including the coupling to the 
$d$-$^3$H ($d$-$^3$He) channels. 
The impact of this coupling can be judged (in the $n+^4$He case) from Fig.~\ref{fig:nHe4_phase_shifts}$(a)$. 
Besides a slight shift of the $P$-wave  
resonances to lower energies,  
the most striking feature is the appearance of a resonance in the $^2D_{3/2}$ partial wave, just above the $d$-$^3$H ($d$-$^3$He) threshold. The further inclusion of distortions of the deuteron 
via an $^2$H $^3S_1$-$^3D_1$ pseudostate $(d^*)$, enhances this resonance, 
while leaving the other partial waves mostly unaffected. In the $d$-$^3$H ($d$-$^3$He) elastic scattering phase shifts of Fig.~\ref{fig:nHe4_phase_shifts}$(b)$ 
we observe a moderate impact of the coupling to the $n$-$^4$He ($p$-$^4$He) channels  
on the $S$-waves. As for the nucleon-$^4$He case, the $^2S_{1/2}$ phase shifts are repulsive due to the Pauli blocking 
(which is treated exactly in our formalism). 
There is, however, a resonance close to threshold in the $^4S_{3/2}$ channel (where projectile and target spins are aligned) that is enhanced by distortions of the deuteron.

\begin{figure}%[t]
\begin{minipage}{8cm}
\includegraphics*[width=0.9\columnwidth]{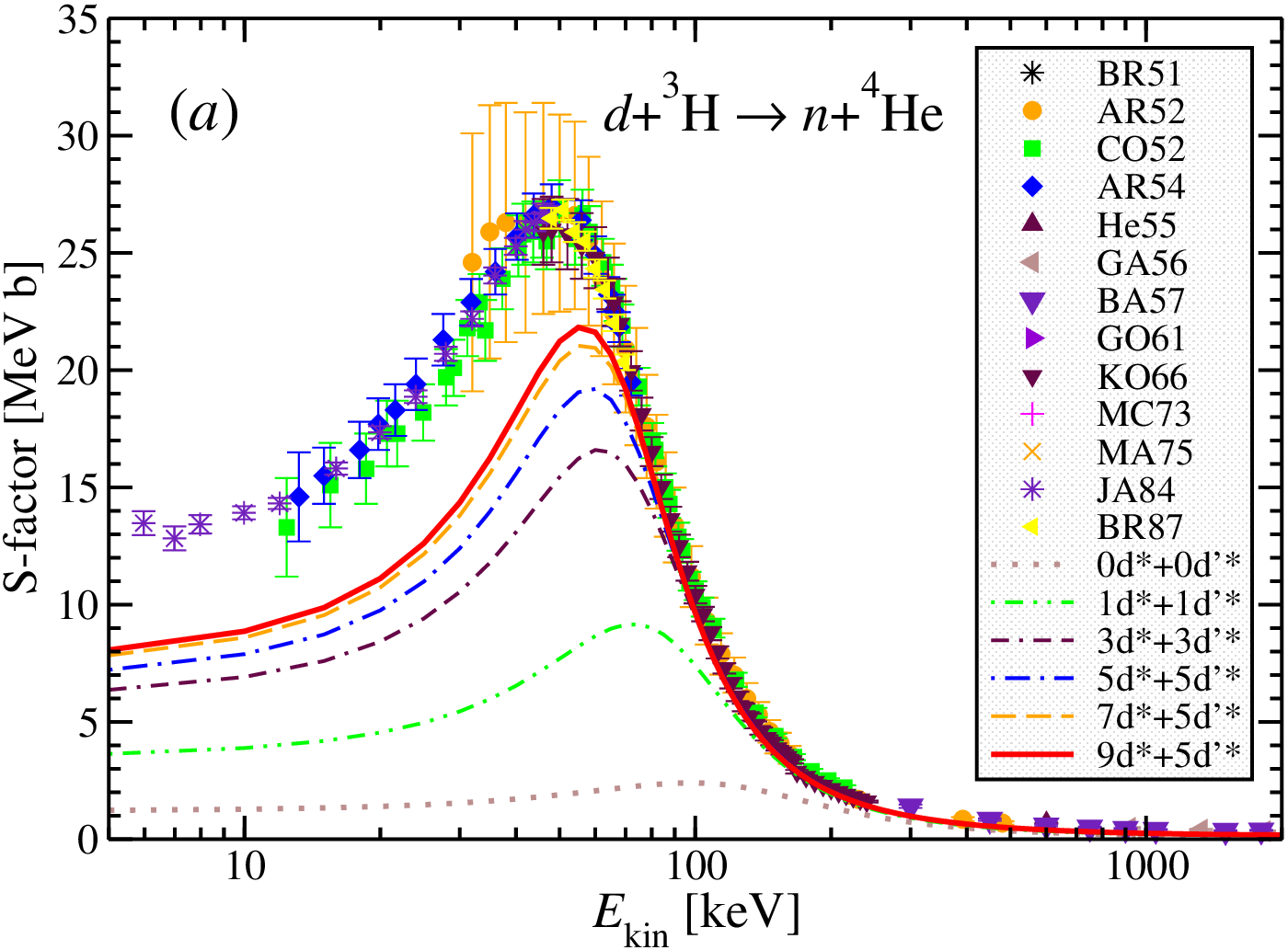}
\end{minipage}
\hfill
\begin{minipage}{8cm}
\includegraphics*[width=0.9\columnwidth]{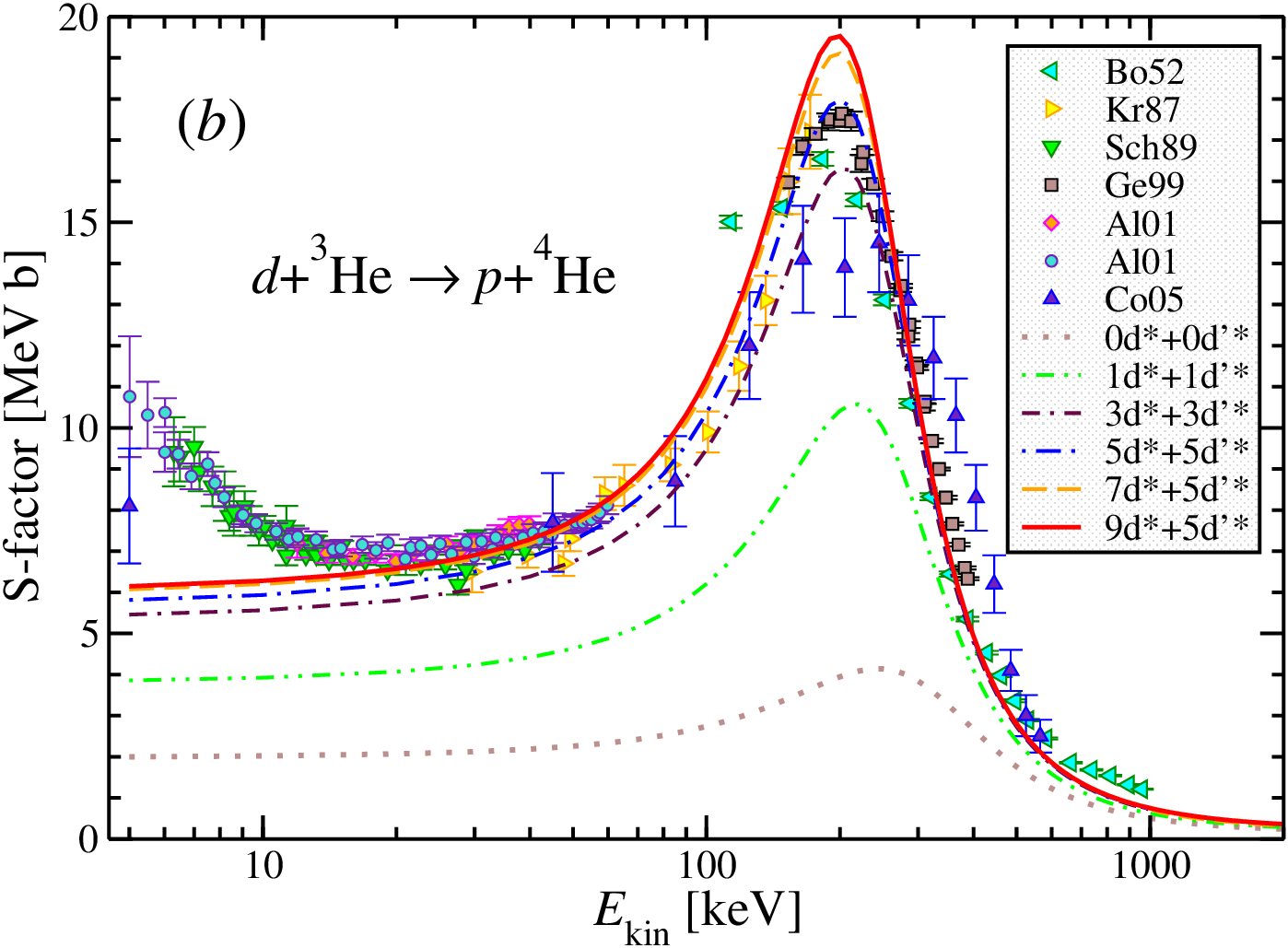}
\end{minipage}
\caption{(Color online) Calculated S-factors of the $^3$H($d$,$n$)$^4$He (a) and $^3$He($d$,$p$)$^4$He (b) reactions compared to experimental data. Convergence with the number of deuteron pseudostates in $^3S_1{-}^3D_1$ ($d^*$) and $^3D_2$ ($d'^*$) channels is shown. See also caption of Fig.~\ref{fig:nHe4_phase_shifts} for details on interaction and HO space used.}
\label{fig:dT_dHe3_Sfact}
\end{figure}

\begin{figure}%[t]
\includegraphics*[width=0.75\columnwidth]{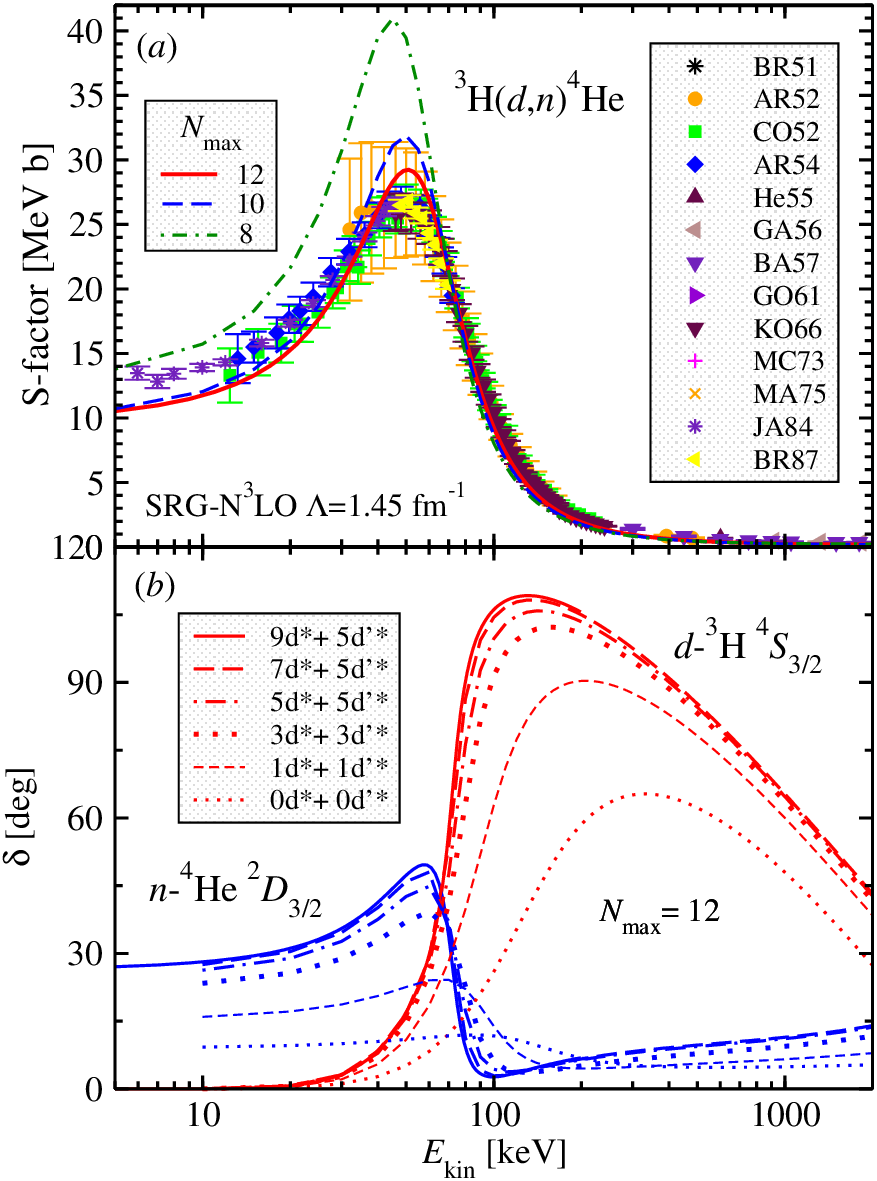}
\caption{(Color online) Calculated  S-factor of the $^3$H($d$,$n$)$^4$He reaction compared to experimental data $(a)$ and diagonal $^2D_{3/2}$ $n$-$^4$He and $^4S_{3/2}$ $d$-$^3$H phase shifts $(b)$. Convergence with $N_{\rm max}$ and the number of deuteron pseudostates in $^3S_1{-}^3D_1$ ($d^*$) and $^3D_2$ ($d'^*$) channels are also shown in $(a)$ and $(b)$, respectively. The $N_{\rm max}{=}8,10,$ and $12$ results contain $9d^*$ plus $3,4,$ and $5d^{\prime*}$, respectively. The $n$-$^4$He kinetic energy is shifted by the $d{-}^3$H threshold energy. The SRG-N$^3$LO $NN$ potential with $\Lambda{=}1.45$ fm$^{-1}$ %, the $N_{\rm max}{=}12$ basis size 
and the HO frequency $\hbar\Omega{=}14$ MeV were used.}
\label{fig:dT_Sfact_1p45}
\end{figure}

Finally, from the scattering phaseshifts we obtain the $^3$H($d$,$n$)$^4$He and $^3$He($d$,$p$)$^4$He cross sections. The corresponding S-factors are  compared to various %experimental 
data sets %~\cite{dT,d3He}
 in panels $(a)$ and $(b)$ of Fig.~\ref{fig:dT_dHe3_Sfact}, respectively.  
The deuteron deformation and its virtual breakup play a crucial role. We show in particular the dependence %of the calculated S-factor 
on the number of $^2$H pseudostates in the $^3S_1$-$^3D_1$ ($d^*$)  and $^3D_2$ ($d^{\prime*}$) channels, included in the calculation. Energies of these pseudostates can be found in Table II of Ref.~\cite{NCSMRGM_dalpha}. The S-factors increase dramatically with the number of pseudostates %included 
until convergence is reached for $9d^*+5d^{\prime*}$. 
Our calculation depends also on the size of the HO basis used to expand the eigenstates of the reacting nuclei as well as the localized parts of the integration kernels (see Eqs.\ (5), (6) and Sec.\ II.\ B.\ of Ref.~\cite{NCSMRGM}). 
As for the bound states, we find a satisfactory convergence [see Fig.~\ref{fig:dT_Sfact_1p45}~$(a)$]. Before demonstrating this point in more detail, here we would like to discuss the comparison with data. 

The experimental position of the  $^3$He($d$,$p$)$^4$He S-factor maximum is well reproduced (within few tens of keV) in our calculations %with the SRG-N$^3$LO $NN$ potential with $\Lambda=1.5$~fm$^{-1}$ 
[Fig.~\ref{fig:dT_dHe3_Sfact} $(b)$]. Overall, the agreement with experiment is quite reasonable, except at very low energies where 
the beam-target data are enhanced by the electron screening. 
For the $^3$H($d$,$n$)$^4$He S-factor, the absolute difference between theoretical and experimental peak positions ($\sim10$ keV) is of the same order of magnitude found in the $d+^3$He case, however the relative difference is much larger for such a low-energy resonance. As a consequence, the $^3$H($d$,$n$)$^4$He S-factor maximum is somewhat underestimated in our calculations and, hence, the calculated S-factor underestimates the data below $\sim 70$ keV. %The $^3$H($d$,$n$)$^4$He and $^3$He($d$,$p$)$^4$He fusion proceed through the $^4S_{3/2}$ resonance in the entrance channel and the $^2D_{3/2}$ resonance in the exit channel. Therefore, a simultaneous accurate description of both these resonances is paramount. 
The inclusion of the $NNN$ force (chiral and SRG-induced) into the calculation should provide closer agreement with experiment, although possibly it would require an $NNN$ interaction accuracy beyond what is currently used in theoretical nuclear physics. However, an even more important consideration is the following. In obtaining the eigenstates of the reacting nuclei, we take into account Coulomb and isospin breaking of the $NN$ interaction. At the same time, we perform isospin projections when evaluating the NCSM/RGM kernels. It is therefore understandable that the splitting between the two peaks may become slightly underestimated in our calculations, so that it is hard to reproduce them equally well simultaneously and a certain amount of tuning of the nuclear interaction may be unavoidable. 

To reproduce the position of the $^3$H($d$,$n$)$^4$He S-factor maximum, we performed additional calculations using SRG-N$^3$LO $NN$ potentials with a lower $\Lambda$.  In Fig.~\ref{fig:dT_Sfact_1p45}, we show that using $\Lambda{=}1.45$~fm$^{-1}$, we are able to reproduce the experimental position of the maximum (we find also a $0.6\%$ variation of the calculated Q value, towards even closer agreement with the measured one). %, without affecting the calculated Q value. 
The theoretical S-factor is then in an overall better agreement with data, although it is slightly narrower and its peak is somewhat overestimated [Fig.~\ref{fig:dT_Sfact_1p45}~$(a)$]. This calculation would suggest that some electron screening enhancement could be also present in the $^3$H($d$,$n$)$^4$He measured S-factor below $\sim 10$ keV.

Finally, the convergence of the calculation with respect to HO basis size and number of deuteron pseudostates is very similar for the two $\Lambda$ values considered. In Fig.~\ref{fig:dT_Sfact_1p45}~$(a)$, we present the $^3$H($d$,$n$)$^4$He S-factor dependence on the size of the HO basis for $N_{\rm max}{=}8{-}12$. We find a satisfactory convergence and expect that an $N_{\rm max}{=}14$ calculation, which is currently out of reach due to computational reasons, would not be significantly different from the present results. 
Also, in Fig.~\ref{fig:dT_Sfact_1p45}~$(b)$ we show the convergence of the $^4S_{3/2}$ and $^2D_{3/2}$ phase shifts with the number of deuteron pseudostates in the vicinity of the $3/2^+$ $^3$H($d$,$n$)$^4$He resonance. This picture is also interesting as it highlights how the $^3$H($d$,$n$)$^4$He and $^3$He($d$,$p$)$^4$He fusion processes proceed through the $^4S_{3/2}$ resonance in the entrance channel and the $^2D_{3/2}$ resonance in the exit channel. The tensor interaction, which is automatically included in the accurate $NN$ potentials we are using, is indispensable for the reaction to take place. %Therefore, a simultaneous accurate description of both these resonances is paramount. 
Unlike the $^4S_{3/2}$, %wave, 
the $^2D_{3/2}$ phase shift does not cross 90 degrees, remaining %. It stays 
positive near the resonance. We note the similarity of our calculated phase shifts with those extracted from the data by using the single-level $R$-matrix fit of Ref.~\cite{Barker97}. In Table~\ref{tab:Sfact}, we summarize our $S(0)$ values and compare them to the $R$-matrix analysis of Ref.~\cite{Desc04}.
\begin{table}[tb]
\caption{Calculated S-factors at zero energy compared to the $R$-matrix data evaluation of Ref.~\protect\cite{Desc04}. The NCSM/RGM calculations as described in Figs.~\ref{fig:dT_Sfact_1p45} and \ref{fig:dT_dHe3_Sfact} for $^3$H($d$,$n$)$^4$He and $^3$He($d$,$p$)$^4$He, respectively.}
\begin{ruledtabular}
\begin{tabular}{ccc}
$S(0)$ [MeV b]            & $^3$H($d$,$n$)$^4$He  & $^3$He($d$,$p$)$^4$He \\
\hline
SRG-N$^3$LO $NN$  &  $10.0 \pm 0.5$                &   $6.0 \pm 0.2$ \\
R-matrix data eval.    &  $11.7 \pm 0.2$             &  $5.9 \pm 0.3$  \\
\end{tabular}
\end{ruledtabular}
\label{tab:Sfact}
\end{table}

In conclusion, we performed {\it ab initio} many-body calculations of the $^3$H($d$,$n$)$^4$He and $^3$He($d$,$p$)$^4$He fusion reactions. Our results are promising and pave the way for microscopic investigations of polarization and electron screening effects, of the $^3$H($d$,$\gamma$)$^5$He radiative capture and other reactions relevant to the fusion research that are less well understood or hard to measure. Our calculations can be further improved by including additional five-body correlations, e.g., virtual breakup of $^3$H ($^3$He). This can be best done by coupling the NCSM/RGM binary-cluster basis with the NCSM calculations for $^5$He ($^5$Li) as outlined in Ref.~\cite{NCSM_review}. Virtual excitations of the deuteron should be treated %The treatment of the deuteron virtual excitations should be best acheived 
by considering explicitly $n$-$p$-$^3$H($^3$He) three-cluster channels. The inclusion of $NNN$ interactions, both chiral and SRG-induced~\cite{JNF09}, is also desirable. Efforts in these directions are under way. 

%\acknowledgments
Computing support for this work came from the LLNL Institutional Computing Grand Challenge program. Prepared in part by LLNL under Contract DE-AC52-07NA27344. Support from the LLNL LDRD grant PLS-09-ERD-020, the U.S.\ DOE/SC/NP (Work Proposal No.\ SCW0498) and the NSERC grant No.\ 401945-2011 is acknowledged.

\end{document}